\documentclass[aps,11pt,notitlepage,nofootinbib,showkeys,prd]{revtex4-1}
\usepackage{amstext,amsmath,amssymb,amsfonts,bbm}
\usepackage[latin1]{inputenc}
\usepackage{epsfig}
\usepackage{hyperref}
\usepackage{amsthm}
\usepackage{subfigure}
\usepackage{color}
\usepackage{multirow}
\newcommand{\bea}{\begin{eqnarray}}	
\newcommand{\eea}{\end{eqnarray}}
\newcommand{\cG}{{\cal G}}

\newcommand{\be}{\begin{equation}}	
\newcommand{\ee}{\end{equation}}

\newcommand{\f}{\frac}
\newcommand{\p}{\partial}

\let\a=\alpha \let\b=\beta  \let\g=\gamma  \let\d=\delta
       \let\k=\kappa 
\let\m=\mu

 \let\D=\Delta

\begin{document}

\title{\Large \bf Phase Transition in Dually Weighted Colored Tensor Models 
}

\author{{\bf Dario Benedetti}}\email{dario.benedetti@aei.mpg.de}
\affiliation{Max Planck Institute for Gravitational Physics (Albert Einstein Institute), \\
Am M\"{u}hlenberg 1, D-14476 Golm, Germany }

\author{{\bf Razvan Gurau}}\email{rgurau@perimeterinstitute.ca}
\affiliation{Perimeter Institute for Theoretical Physics, 31 Caroline St. N, ON N2L 2Y5, Waterloo, Canada}

\begin{abstract}

Tensor models are a generalization of matrix models (their graphs being dual to higher-dimensional triangulations)
and, in their colored version, admit a  $1/N$ expansion and a continuum limit.
We introduce a new class of colored tensor models with a modified propagator which allows us to associate weight 
factors to the faces of the graphs, {\it i.e.} to the bones (or hinges) of the triangulation, 
where curvature is concentrated. They correspond to dynamical triangulations in three and higher dimensions 
with generalized amplitudes. We solve analytically the leading order in $1/N$ of the most general model in arbitrary dimensions. 
We then show that a particular model, corresponding to dynamical triangulations with a non-trivial measure factor, 
undergoes a third-order phase transition in the continuum characterized by a jump in the susceptibility exponent.
\end{abstract}

\medskip

\begin{flushright} \small
AEI-2011-057\\
PI-QG-235
\end{flushright}
\noindent  Pacs numbers: 02.10.Ox, 04.60.Gw, 04.60.Nc
\keywords{1/N expansion of random tensor models, critical behavior, dynamical triangulation}

\maketitle

\section{Introduction}

Statistical models of fluctuating geometry are a generous source of results and ideas for physics and mathematics.
A particularly attractive feature of many such models is that they can be thought as providing either a regularization 
or a fundamental description of quantum gravity \cite{ambjorn-book}.
Dynamical Triangulations (DT) \cite{david-revueDT,ambjorn-houches94} are one of the most studied examples, and have been very 
successful in two dimensions, where they are related to the large-$N$ limit of matrix models \cite{Kaz,Dav}, and whose link to non-critical 
string theory in the continuum limit is well-understood \cite{Di Francesco:1993nw}. 

Higher-dimensional models of DT have not been equally successful in providing a sensible continuum limit for quantum 
gravity \cite{Loll:1998aj}, leading to degenerate geometries at large scales \cite{Ambjorn:1995dj} with a first-order phase transition separating them \cite{Bialas:1996wu,deBakker:1996zx}.
However one cannot exclude that some unknown essential ingredient was missing in the models analyzed so far. 
A non-local modification of DT, which goes under the name of causal dynamical triangulations or CDT \cite{Ambjorn:2010ce} has produced 
substantial evidence for the emergence of an extended geometry at large scale \cite{Ambjorn:2005qt,Ambjorn:2011ph}, 
and signs of a second-order phase transition \cite{Ambjorn:2011cg},
hinting at the possibility to attain a good continuum limit. 
Recently it has also been suggested \cite{Laiho:2011ya} that the effects of a non-trivial
measure factor were overlooked in the past and could potentially lead to an improvement of the large scale behavior of DT models.

Tensor models \cite{ambj3dqg,mmgravity,sasa1} and group field theories \cite{danielegft}
are the generalization of matrix models to higher dimensions. In particular, the colored tensor models and group field theories 
\cite{color,lost,PolyColor} generate graphs dual to orientable \cite{orie} pseudo-manifolds in any dimension. 
Much progress has been done in understanding these models. 
Various power counting estimates and bounds of graph amplitudes in tensor models
and group field theories have been obtained \cite{pcont1,pcont2,pcont3,pcont4,pcont5,pcont6,pcont7}. The symmetries of tensor models have
been analyzed either with the help of n-ary algebras \cite{sym1,sym2,sym3}, or,  in a more quantum field theoretical 
approach through Ward-Takahashi identities \cite{sym4}. The relation between symmetries of group field theories and the
diffeomorphism symmetry of the resulting triangulation has been explored \cite{sym5,sym6}. Solutions of the classical 
equations of motions \cite{class1,class2} have been derived and some of them interpreted as matter fields on non 
commutative spaces.
Most importantly, for our purposes, the colored tensor models have been shown to possess an additional (and welcomed) feature as compared to non-colored models: their amplitudes are such that a $1/N$ expansion 
is possible, with the leading order encoding a sum over a class of colored triangulations of the $D$-sphere \cite{Gur3,GurRiv,Gur4}.
This discovery led to the possibility of new analytical  investigations of DT models and their continuum limit 
in $D\geq 3$ dimensions \cite{Bonzom:2011zz,Gurau:2011tj}.

The link between tensor models and DT leads straight to a daunting question: will any colored tensor model admit a richer continuum limit? 
The question is twofold, as at first instance one can wonder whether the coloring will 
already suffice to generate new universality classes in these models. The results of  \cite{Bonzom:2011zz} indicate that, as far as the 
critical exponents are concerned, the color alone is not enough and the continuum limit is strongly reminiscent of branched 
polymers\footnote{The reader should keep in mind however that no other characteristic of branched polymers, like say the Hausdorff 
or spectral dimension has so far been reproduced for the colored tensor models.}. 
The next question is then whether it is possible to modify the tensor models in such a way that new, appealing, phases 
would appear (like in CDT) and/or a second-order phase transition would occur between phases. 

The dually weighted matrix models introduced by Kazakov et al. \cite{Kazakov:1995ae} are an ideal candidate for 
introducing relevant modifications of a two-dimensional DT model via a local modification of the corresponding 
matrix model (see also \cite{Kazakov:1995gm,Kazakov:1996zm,Szabo:1996fj}). 
For example, one can impose the non-local condition on foliations that characterizes two-dimensional CDT 
just by a suitable modification of the propagator of a two-matrices model \cite{Benedetti:2008hc}. 
 
In the present work we extend the idea of dually weighted matrix models to colored tensor models (which by lack of fantasy we call 
dually weighted colored tensor models) and solve them analytically at leading order in $1/N$.
We then consider a particular example corresponding to a DT model initially proposed in \cite{Bilke:1998vj} and recently 
claimed  to exhibit a new phase with promising geometrical properties \cite{Laiho:2011ya}. In the large-$N$ limit, 
equivalent to weak-coupling from the gravitational point of view, we find a third-order phase transition 
and a new critical behavior. We compute explicitly the susceptibility exponent associated to the different phases. 
Although, as we are in a different regime\footnote{And have not computed the spectral dimension of the emerging geometry.}, 
we cannot make direct contact to the results of  \cite{Laiho:2011ya}, our results show that a non-trivial continuum limit is 
possible in these models and open a small window of hope on non-causal DT models.

This paper is organized as follows. In Sec.~\ref{Sec:dwctm} we introduce the new models, discuss their DT interpretation and 
derive a Schwinger-Dyson equation relating the free energy and the connected two-point function of our new tensor models. 
In Sec.~\ref{Sec:largeN} we derive, in the large-$N$ limit, a set of self-consistency equations which allow to exactly 
solve (at leading order in $1/N$) the model. In Sec.~\ref{Sec:beta} we specialize to a particular model and study its phase transition.

\section{Dually weighted colored tensor models}
\label{Sec:dwctm}

In this section we introduce a modification of the independent identically distributed (i.i.d.) 
colored tensor model\footnote{The epithet i.i.d. refers to the fact that in such models the free Gaussian measure weights independently and identically each component of each tensor. In the new models we are introducing this is not the case.} of \cite{color,lost,PolyColor,Gur3,GurRiv,Gur4,Bonzom:2011zz}, which we baptize
``dually weighted colored tensor model''.

\begin{figure}[t]
 \includegraphics[width=6cm]{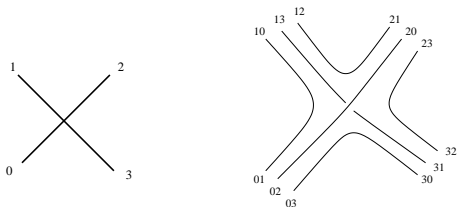}
  \caption{The index structure of a $D=3$ vertex, in a ``coarse grained'' view, showing only the tensor color, and in a detailed view,
 showing the strands (indices should be read clockwise).}
\label{fig:vertex}
\end{figure}

We denote $\vec n_i$, for $i=0,\dotsc,D$, the $D$-tuple of integers
 $\vec n_i = (n_{ii-1},\dotsc, n_{i0},\; n_{iD}, \dotsc, n_{ii+1}) $, with
$n_{ik}=1,\dotsc, N$. This $N$ is the size of the tensors and the large $N$ limit defined in 
\cite{Gur3,GurRiv,Gur4} represents the limit of infinite size tensors. 
We set $n_{ij} = n_{ji}$. Let $\bar \psi^i_{   \bar {\vec n}_i },\; \psi^i_{\vec n_i}$, with $i=0,\dotsc, D$, be $D+1$ couples of complex
conjugated tensors with $D$ indices\footnote{We denote the indices of the $\bar \psi$ field by $\bar n$. This does not denote a complex 
conjugation, being merely a book keeping device.}. The dually weighted colored tensor model in dimension $D$ is defined by the partition 
function
\bea
 e^{ N^D E } &=& Z_N(\lambda, \bar{\lambda}) = \int \, d\bar \psi \, d \psi
\ e^{- S (\psi,\bar\psi)} \; , \crcr
 S (\psi,\bar\psi) &=& \sum_{i=0}^{D} \sum_{ {\vec p}_i, \bar {\vec n}_i } \psi^i_{ {\vec p}_i} \Big{(} 
\prod_{j} (C^{-1})_{ p_{ij}  \bar n_{ij}  } \Big{)} \bar \psi^i_{\bar {\vec  n}_i} 
\crcr
&+&
\frac{\lambda}{ N^{D(D-1)/4} } \sum_{ n} \prod_{i=0}^D \psi^i_{ \vec n_i } +
\frac{\bar \lambda}{ N^{D(D-1)/4} } \sum_{ \bar n}
\prod_{i=0}^D \bar \psi^i_{ \bar {\vec n}_i } \; . \label{eq:EDT}
\eea 
$\sum_{ n}$ denotes the sum over all indices $ n_{ij}$ from $1$ to $N$. 
The index structure of a vertex is represented in Fig.~\ref{fig:vertex}.
The quadratic part is chosen such that Gaussian correlations, i.e. the propagator of
the model, is (no summation over $i$, the propagator between fields of different color is zero)
\bea 
 \langle  \bar \psi^i_{ \bar {\vec  n}_i} \psi^i_{ {\vec p}_i}   \rangle_0 = \prod_{j} C_{ \bar n_{ij}  p_{ij} } \, ,
\qquad 
 \langle   \psi^i_{{\vec p}_i } \bar \psi^i_{ \bar {\vec  n}_i}   \rangle_0 =  \prod_{j} C_{ \bar n_{ij}  p_{ij} } =  \prod_{j} C^T_{  p_{ij}  \bar n_{ij}}
\; ,
\eea 
where $C^T$ denotes the transposed matrix. Obviously, when $C_{ \bar n_{ij}  p_{ij} }=\d_{ \bar n_{ij}  p_{ij} }$ the model reduces to the 
i.i.d. one. 

\begin{figure}[t]
 \includegraphics[width=3cm]{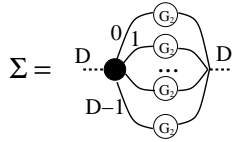}
  \caption{The self energy $\Sigma$ at leading order in terms of the connected two-point function $G_2$. The labels $0,1,\dots D$ 
    denote the colors of various lines.}
\label{fig:twopoint}
\end{figure}

Graphs are made of vertices, (colored) lines and ``faces''. Like in matrix models, the indices of the 
tensors are associated to ``strands'' (the solid line in the detailed view of the vertex in figure \ref{fig:vertex}).
A index $n_{ij}$ (associated to the strand common to the half lines of color $i$ and $j$ of the vertex) 
is identified by the two colors, $i$ and $j$. The colors are conserved along the lines, hence the ``faces'' 
(closed strands) of the graph are identified by couples of colors. A graph is dual to a triangulation\footnote{More precisely to an abstract finite 
simplicial pseudo-manifold \cite{lost}.}: its vertices are dual 
to $D$ simplices, its lines to $D-1$ simplices and its faces to $D-2$ simplices.
When computing amplitudes of graphs, one picks 
up the trace of the alternate product of
$C$ and $C^T$ along each face of a graph. We consider only matrices $C$ which do not modify the scaling in $1/N$ of the amplitude, 
{\it i.e.} such that $ \lim_{N\to \infty} \frac{\text{Tr}[(CC^T)^q]}{N}$ is finite for all $q$.
We {\it do not} impose for the moment any further restrictions on $C$, in particular $C$ is not required to be hermitian.
Using the counting of faces of a colored graph established in \cite{Gur4}, the amplitude writes
\bea
  A(\cG) = (\lambda \bar \lambda)^p \Big{(} \prod_{(ij), \rho} 
\frac{\text{Tr}[ (CC^T)^{p^{ij}_{(\rho)} }]}{N} \Big{)} \;
N^{ D - \frac{2}{(D-1)!} \omega(\cG) } \; ,
\eea 
with $2 p^{ij}_{(\rho)}$ the number of vertices of the $\rho$'th face of colors $ij$.
The number $\omega(\cG) \ge 0$ is called the degree of the graph $\cG$ \cite{Gur4}, and the leading order graphs are those 
of degree $0$, which are dual to triangulations of $D$-spheres \cite{GurRiv}. Crucially for the study of the leading sector in the large $N$ limit is the flowing fact \cite{Bonzom:2011zz}: 
at leading order in $1/N$ the self energy $\Sigma$ ({\it i.e.} one particle irreducible amputated two-point function) factors 
into the convolution of $D$ connected two-point functions $G_2$, one for each color. Such graphs (represented schematically 
in Fig.~\ref{fig:twopoint}) are called 
melons\footnote{This is stark contrast with the $D=2$ case of usual matrix models. Indeed, the $1/N$ expansion
in that case selects all planar graphs, not only the melonic ones. As shown in \cite{Bonzom:2011zz} the results leading to the
factorization of the self energy explicitly break down in $D<3$.}. 
Denoting $g=\lambda\bar\lambda$,
the free energy of the model is a function $E( C_{  \bar   p   n } ,g)$ of
the coupling $g$ and the matrix entries $ C_{  \bar  p   n }$.

For $C=\mathbb{I}$  the i.i.d. model is in direct correspondence with DT in $D$ dimensions, just like matrix models are with 
DT in two dimensions. Each graph is dual to a triangulation, 
and denoting $N_k$ the number of $k$-simplices, its amplitude rewrites 
\be\label{eq:ampligraph}
A(\cG) = e^{\kappa_{D-2}N_{D-2}\, -\, \kappa_D N_D}\, ,
\ee
where $\kappa_{D-2}$ and $\kappa_D$ are (with $g=\lambda\bar{\lambda}$),
\be \label{DTcouplings}
\kappa_{D-2} = \ln N,\quad \text{and} \qquad
\kappa_D = \frac12\Bigl( \frac12\,D(D-1)\, \ln N - \ln (g)\Bigr)\, .
\ee
The grand canonical partition function of DT is given by
\be
Z_{DT} = \sum_T 	\f{1}{s(T)} e^{\kappa_{D-2}N_{D-2}\, -\, \kappa_D N_D} \, ,
\ee
where the sum is over all $D$-dimensional triangulations $T$, and $s(T)$ is the order of the automorphism group of $T$.
The grand canonical partition function of DT equals the free energy of the tensor model, $Z_{DT}=E$.

The large-$N$ limit corresponds to $\k_{D-2}\to \infty$, with $\k_D = \tfrac14 D(D-1) \k_{D-2} - \tfrac12\ln g$. As the DT action can be 
interpreted as a Regge action for equilateral simplices, one has $\k_{D-2}\sim 1/G$ thus the large-$N$ limit amounts to the weak coupling 
limit, with vanishing (bare) Newton's constant. Note that this is true also for matrix models, as in two dimensions the Regge action just 
gives the Euler character of the surface, and the limit $G\to 0$ is the planar limit. As in that case, in order to keep a finite $G$ one 
would have to perform some kind of double scaling limit.

For $C\neq \mathbb{I}$ the amplitudes associated to triangulations will be in general different, thus defining new models of DT. In particular, 
the effect of the modified propagator is to associate a weight factor to each $(D-2)$-subsimplex of the triangulation, also called a bone (or hinge) 
in Regge calculus. As in Regge's construction it is precisely on the bones that curvature resides, the dually weighted colored tensor models can be a
precious tool for the study of DT with more a complicated curvature dependence of the action.

\subsection{The connected two-point function}

The full connected two-point function of the dually weighted model,
\bea
 \langle  \bar \psi^i_{ \bar {\vec  n}_i} \psi^i_{ {\vec p}_i}   \rangle    
 =  \prod_{j} P_{  \bar n_{ij}  p_{ij}  }(g,C) \; , 
\eea 
is of course independent of the color $i$ and factored along strands. More subtly, the 
contribution of the strand of color $ij$ to the two-point function is in fact independent of $j$. 
This is due to the fact that the action is invariant under a transformation which permutes any two 
strands $ij$ and $ik$ in the field $\psi^i$ and permutes the strands on all other fields (and also the fields between
themselves) to restore the connectivity of the vertex.

We are now going to show that knowledge of the full connected two-point function suffices to compute the derivatives of the free energy. First, 
we observe that from
\bea
\frac{1}{Z} \sum_{ \bar {\vec n}_i }  \int \frac{\delta}{\delta \bar \psi_{ \bar {\vec  n}_i } } 
 \Big{(} \bar \psi_{ \bar {\vec  n}_i } e^{-S}  \Big{)} =0 \;,
\eea 
it follows that
\bea
 N^D -   \sum_{ \bar {\vec n}_i , \vec p_i }   \prod_{j} (C^{-1})_{ p_{ij}  \bar n_{ij}  }  
\langle  \bar \psi^i_{ \bar {\vec  n}_i} \psi^i_{ {\vec p}_i}   \rangle + \frac{1}{Z} \bar \lambda
 \partial_{\bar \lambda} Z =  N^D -  \prod_{j} \text{Tr}[C^{-1}P] + \bar \lambda N^D
 \partial_{\bar \lambda} E =0 \; , \crcr
\eea 
which, recalling that $E$ is a function only of $g$, can be rewritten as
\bea
 \bar \lambda \partial_{\bar \lambda} E = g\partial_g E =  \Big{[} \frac{   \text{Tr}[C^{-1}P] }{N} \Big{]}^{D} -1 \; .
\eea 
Furthermore, we have
\bea
 \frac{\partial E}{\partial C_{ \bar n  p }} = - N^{-D} \sum_{ {\vec a}_i, \bar {\vec b}_i } 
\Bigl( \prod_{j} \frac{\partial (C^{-1})_{ a_{ij}  \bar b_{ij}  } }{ \partial C_{ \bar  n   p  } }  \Bigr)
\Big{\langle}  \psi^i_{ {\vec a}_i} \bar \psi^i_{\bar {\vec  b}_i} 
 \Big{\rangle} \; ,
\eea 
which, using $ \partial_{ C_{ \bar  n  p  } } (C_{\bar c a } C^{-1}_{a \bar b} )=
\delta_{ \bar  n \bar c  } C^{-1}_{p \bar b} + C_{\bar c a } \partial_{  C_{ \bar  n  p  }  } C^{-1}_{a \bar b} =0
$, yields 
\bea
&& \frac{\partial E}{\partial C_{ \bar n  p }} = N^{-D} \sum_{ {\vec a}_i, \bar {\vec b}_i } 
\Bigl( \prod_{j} 
\bigl(C^{-1}_{a_{ij} \bar n}   C^{-1}_{p \bar b_{ij}} \bigr) \Bigr)
\Big{\langle}  \psi^i_{ {\vec a}_i} \bar \psi^i_{\bar {\vec  b}_i} 
 \Big{\rangle} 
 =N^{-D} \prod_{j} (C^{-1}PC^{-1})_{p\bar n} \crcr
&&=  \Bigl(  \frac{   (C^{-1}PC^{-1})_{p\bar n}  }{N}\Bigr)^D \; .
\eea 
Solving the model consists therefore in determining $P$. We will do this in the next section.

\section{The leading order in $1/N$}
\label{Sec:largeN}

At leading order only melonic two-point functions contribute \cite{Bonzom:2011zz}. They are characterized, as we already mentioned, by the fact 
that the self energy of the model $\Sigma= \langle  \bar \psi^i_{ \bar {\vec  n}_i} \psi^i_{ {\vec p}_i}   \rangle_{1PI, amputated}$ 
is given by the convolution (respecting the strand structure) of connected two-point
functions. Supplementing this by the classical Schwinger-Dyson equation relating the full two-point function $G_2$, the self energy
$\Sigma$ and the propagator $C$ of a field theory
\bea
  G_2 = C \frac{1}{1-\Sigma C} \; ,
\eea 
yields the system of equations
\bea
&& \sum_{\vec q_{i}   } \Bigl[ \langle  \bar \psi^i_{ \bar {\vec  n}_i} \psi^i_{ {\vec q}_i}   \rangle 
\Big{(} \prod_j \delta_{ q_{ij}  p_{ij}  } - \sum_{\bar {\vec r}_{i}   }
\langle  \psi^i_{ {\vec q}_i}    \bar \psi^i_{ \bar {\vec  r}_i}  \rangle_{1PI, amputated}
   \prod_{j} C_{  \bar r_{ij}  p_{ij}  } \Big{)} \Bigr]=  \prod_{j} C_{  \bar n_{ij}  p_{ij}  } \; ,\crcr
&&\langle  \psi^i_{ {\vec q}_i}    \bar \psi^i_{ \bar {\vec  r}_i}  \rangle_{1PI, amputated} = 
g N^{-\frac{D(D-1)}{2} } \sum_{{\vec q}_j,  \bar {\vec  r}_j \neq  {\vec q}_i,\bar {\vec  r}_i}   \prod_{j\neq i} 
   \langle  \psi^j_{ {\vec q}_j}    \bar \psi^j_{ \bar {\vec  r}_j}  \rangle \; .
\eea 
Substituting the connected two-point function and performing the sums leads to 
\bea\label{eq:fundamental}
 \prod_{j} P_{  \bar n_{j}  p_{j}  } - g  N^{-\frac{D(D-1)}{2} } [\text{Tr}(PP^{T})]^{ \frac{D(D-1)}{2} }
   \prod_{j} ( PP^{T} C)_{  \bar n_{j}  p_{j}  }
   = \prod_{j} C_{  \bar n_{j}  p_{j}  } \; ,
\eea 
where 
 the index $i$ has been erased as it plays no role. Surprising as it might be, 
equation \eqref{eq:fundamental} can be solved analytically. To do so, we first 
introduce a matrix $X =C^{-1}P$, 
we multiply by $X_{p_{j} b_{j}}$ from the right and by $P^{-1}_{ a_{j} \bar n_{j} }$ from the left. Summing the resulting expression over all $\bar n_{j}$ and $p_{j}$ (with range 
from $1$ to $N$) we get
\bea
  \prod_{j} X_{a_{j} b_{j}} = \prod_j \delta_{a_{j} b_{j} }
+ g N^{-\frac{D(D-1)}{2} } [\text{Tr}(PP^{T})]^{ \frac{D(D-1)}{2} }  \prod_{j} ( P^{T} P )_{  a_{j}  b_{j}  }  \;.
\eea 
To solve for $X$, we first take $a_j=b_j$ for all $j$ and sum, obtaining
\bea
 [\text{Tr} (X)]^D = N^D + g N^{-\frac{D(D-1)}{2} } [\text{Tr}(PP^{T})]^{ \frac{D(D+1)}{2} } \; ,
\eea 
and then we take $a_j=b_j$ for all but one $j$ and sum,  obtaining
\bea
 X_{ab} [\text{Tr} (X)]^{D-1} = \delta_{ab} N^{D-1} + g N^{-\frac{D(D-1)}{2} } [\text{Tr}(PP^{T})]^{ \frac{D(D+1)}{2} -1 } 
   (P^T P)_{ab} \; .
\eea 
Combining the two equations we have
\be
X = 
\frac{ N^{D-1}  \mathbb{I} +  g N^{-\frac{D(D-1)}{2} } [\text{Tr}(PP^{T})]^{ \frac{D(D+1)}{2} -1 }    P^T P  }
{ \Big{(} N^D + g N^{-\frac{D(D-1)}{2} } [\text{Tr}(PP^{T})]^{ \frac{D(D+1)}{2} }  \Big{)}^{\frac{D-1}{D}} } \; ,
\ee
and we finally get the following expression for $C$ as a function of $P$,
\bea\label{eq:CP}
 C = P \frac{ \bigl( 1+g\alpha^{ D(D+1) } \bigr)^{\frac{D-1}{D}} } 
 {  \mathbb{I} + g \alpha^{ D(D+1)-2  } P^T P } \, ,
\qquad  C^TC = \frac{ \bigl( 1+g\alpha^{ D(D+1)  } \bigr)^{2 \frac{D-1}{D}} }
{ \Bigl( \mathbb{I} + g \alpha^{D(D+1)-2 } P^T P   \Bigr)^2 } P^T P \; ,
\eea 
where  $\alpha^2 = \tfrac{1}{N} \text{Tr}(PP^T)=\tfrac{1}{N} \text{Tr}(P^T P)  $.
As $C^TC$ is a function of  $P^T P$, the two commute, and equation \eqref{eq:CP} can be written
as a quadratic equation for $P^T P$ in terms of $C^TC$, 
whose physical solution is obtained by choosing the sign of the root that gives $P^T P=0$ when $C^TC=0$:
\bea
 P^T P = \frac{1}{  2 g^2 \alpha^{ 2D(D+1)-4} C^TC  } \Bigg{[} &&
   \bigl( 1+g\alpha^{  D(D+1)  } \bigr)^{2 \frac{D-1}{D}} \mathbb{I}
- 2 g \alpha^{ D(D+1)-2 } C^TC  \crcr
&&-  \bigl( 1+g \alpha^{ D(D+1)  } \bigr)^{2 \frac{D-1}{D}} 
  \sqrt{ \mathbb{I} -  \frac{4 g \alpha^{D(D+1)-2 }   } { 
 \bigl( 1+g \alpha^{ D(D+1)  } \bigr)^{2 \frac{D-1}{D}}   }
    C^TC  }\Bigg{]} \; ,
\eea 
Finally, expanding the square root in Taylor series we can write
\bea\label{eq:sol}
&& P^T P =  \sum_{q=1} \frac{1}{q+1} \binom{2q}{q} 
\frac{ \Big{[}g \alpha^{D(D+1)-2  } \Big{]}^{q-1}  }
{ \Big{[} \bigl( 1+g \alpha^{ D(D+1)  } \bigr)^{2 \frac{D-1}{D}} \Big{]}^q  } 
    (C^TC)^q  \; ,\\ 
\label{eq:solsc}
&& \alpha^2 = \sum_{q=1} \frac{1}{q+1} \binom{2q}{q} 
\frac{ \Big{[}g \alpha^{D(D+1)-2  } \Big{]}^{q-1}  }
{ \Big{[} \bigl( 1+g \alpha^{ D(D+1)  } \bigr)^{2 \frac{D-1}{D}} \Big{]}^q  } 
    \frac{\text{Tr}[(C^TC)^q]}{N}  \; .
\eea 
Equation \eqref{eq:sol} determines $P^T P $ in terms of $ C^TC $ and 
$\alpha$, whereas \eqref{eq:solsc} implicitly defines $\alpha$ in terms of the traces of powers
of $ C^TC $ yielding the analytic solution at leading order in $1/N$ of the dually weighted colored tensor model.
Note that the derivative of the free energy with respect to $g$ is
\bea
g\partial_g  E  =  \Big{[} \frac{1}{N} \text{Tr}[C^{-1}P] \Big{]}^D - 1= 
\Big{(}  1 + g N^{-\frac{D(D+1)}{2} } [\text{Tr}(PP^{T})]^{ \frac{D(D+1)}{2} }  \Big{)} -1
 = g\alpha^{ D(D+1) } \; ,
\eea 
hence studying only the self-consistency equation \eqref{eq:solsc} for $\alpha$ suffices to study the critical behavior
of the model.
For $C=\mathbb{I}$ a straightforward computation leads from \eqref{eq:solsc} to $\a^D = 1+g\a^{D(D+1)}$, 
reproducing the solution of the i.i.d. model found in \cite{Bonzom:2011zz}. 

\section{Phase portrait of a particular model}
\label{Sec:beta}

In order to draw the phase portrait of a specific model it is more convenient to denote 
$g\partial_gE = g \alpha^{D(D+1)} \equiv U$ and write the self-consistency equation \eqref{eq:solsc} in terms of $U$ as
\bea \label{selfceq}
 U = \sum_{q=1} \frac{1}{(q+1) } \binom{2q}{q} 
   \Big{[} \frac{ U^{1-\frac{2}{D(D+1) } }   }
{  \bigl( 1+ U \bigr)^{2 \frac{D-1}{D}}  }   g^{\frac{2}{D(D+1)}}  \Big{]}^{q}
  \;  \frac{\text{Tr}[(C^TC)^q]}{N} \; .
\eea 

In the remainder of this paper we deal with the model defined by a covariance $C$ such that 
\bea \label{betaC}
 \text{Tr}[ (C^TC)^q] = N q^{-\beta} \; .
\eea 
Such a choice corresponds to the DT amplitude
\be\label{eq:ampliC}
A(\cG) = e^{\kappa_{D-2}N_{D-2}\, -\, \kappa_D N_D} \prod_{i} q_i^{-\b} \, ,
\ee
where the product is over all $(D-2)$-dimensional simplices (bones) of the triangulation, with $q_i$ being the number of
 $D$-simplices to which the bone $i$ belongs.
The DT amplitude \eqref{eq:ampliC} was studied via numerical simulations in \cite{Bilke:1998vj}, and more recently in \cite{Laiho:2011ya}, 
where it was argued that a new phase  appears, for large enough $\b$, with promising geometrical properties.
We will now solve analytically in the large-$N$ limit the corresponding dually weighted colored tensor model 
defined by \eqref{betaC} .

First note that a matrix $C$ satisfying \eqref{betaC} exists in the large-$N$ limit. 
To determine $C$ we can for instance diagonalize it 
and write $N$ equations for the $N$ eigenvalues,
corresponding to the traces up to power $N$. 
These equations will always have roots in the complex domain\footnote{Actually it can be shown that for given $N$ the solution is unique up to permutations of the eigenvalues.}, thus we obtain in general a non-hermitian matrix $C$ which satisfies \eqref{betaC} up to $q=N$.
Alternatively we can impose a weaker condition $\text{Tr}[ (C^TC)^q] = N q^{-\beta} + O(1)$ for every $q\geq 1$, and solve for the eigenvalue distribution in the large-$N$ limit, using standard techniques from matrix models. This way the spectrum of $C^TC$ can be chosen to be real.

The self-consistency equation \eqref{selfceq} (and its physical initial condition) is now
\bea\label{eq:sys}
&&  U = S\bigl(\beta, z(g,U) \bigr) \; ,  \qquad 
 S(\beta,z) =  \sum_{q=1}^\infty \frac{1}{q^{\beta} (q+1) } \binom{2q}{q} z^q \crcr
&&z(g,U) =  \frac{ U^{1-\frac{2}{D(D+1) } }   }
{  \bigl( 1+ U \bigr)^{2 \frac{D-1}{D}}  }   g^{\frac{2}{D(D+1)}} \; , \qquad 
g( \beta, 0  ) =0 \; .
\eea 
We will denote $g(\beta, U)$ the solution of the equation \eqref{eq:sys}.

\begin{figure}
\begin{center}
$\begin{array}{c@{\hspace{1in}}c}
\multicolumn{1}{l}{} &
    \multicolumn{1}{l}{} \\ [-0.53cm]
\includegraphics[width=6.5cm]{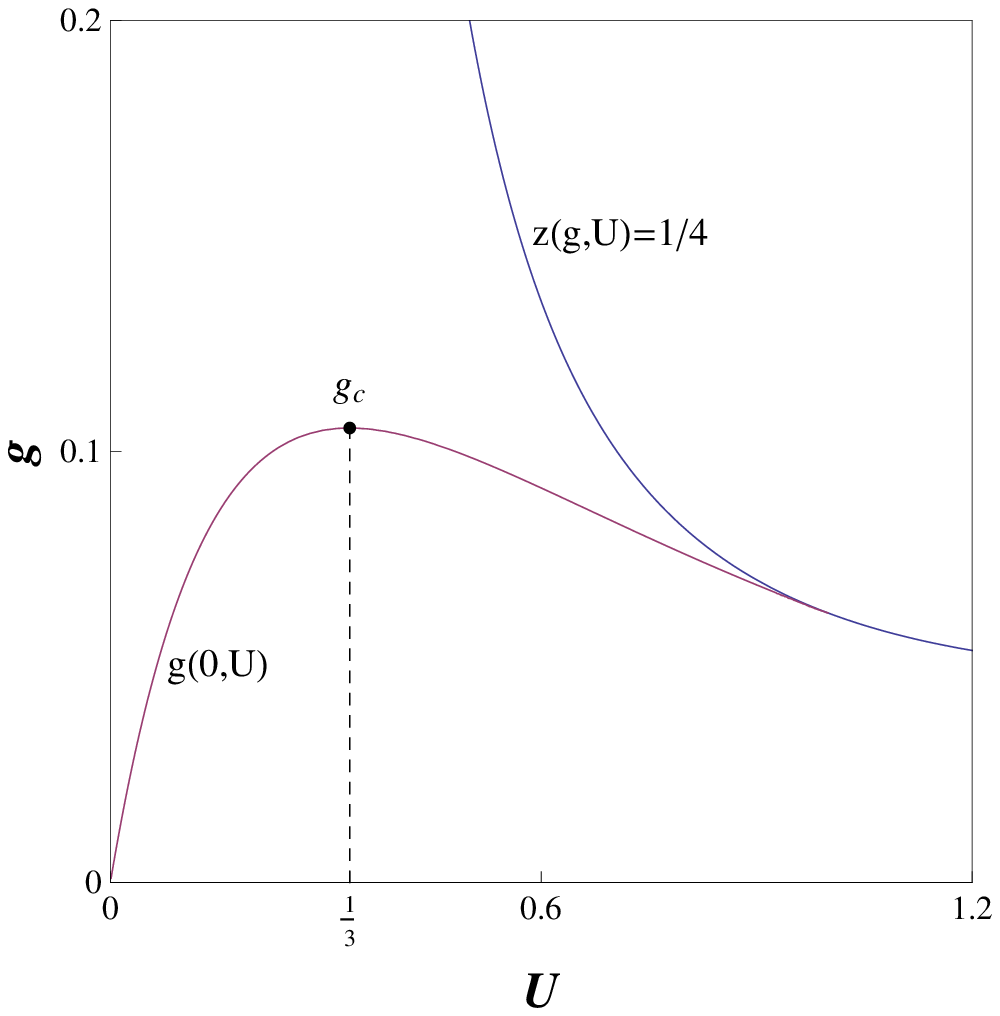} &
\includegraphics[width=6.5cm]{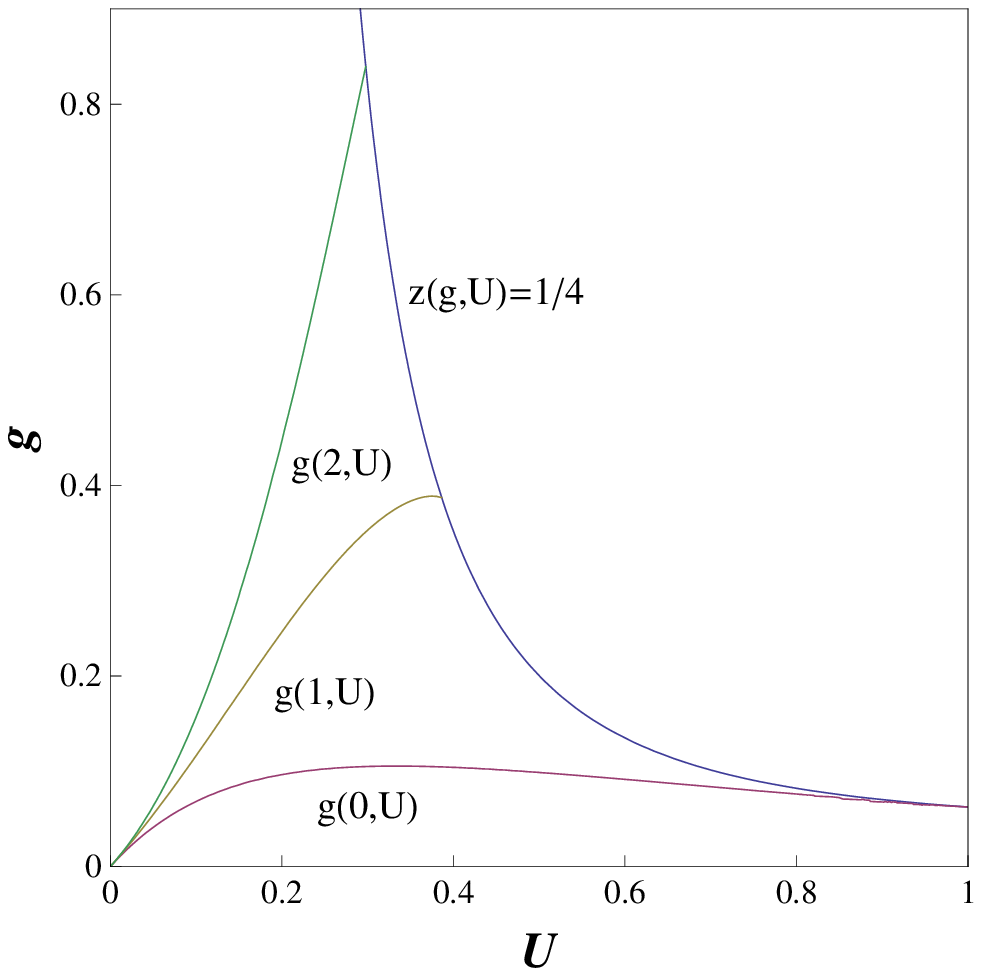} \\ [0.4cm]
\mbox{\bf (a)} & \mbox{\bf (b)}
\end{array}$
\end{center}
\caption{
The curve $ g(\b, U)$, for $D=3$ and for $\b=0$ \mbox{\bf (a)} and  $\b=0,1,2$ \mbox{\bf (b)}, together with the critical curve 
$z\bigl(g_b(U),U \bigr)=1/4$. 
}
\label{fig:curves}
\end{figure}

The series $S(\b,z)$ has radius of convergence $z_b=\frac{1}{4}$ for all values of $\beta$, that is, it converges for all 
$g$ and $U$ under the curve
\bea  \label{gb}
z \bigl(g_b(U),U \bigr)=\frac{1}{4} \Rightarrow g_b(U)= \frac{ (1+U)^{D^2-1}}{ 4^{ \frac{D(D+1)}{2} } U^{ \frac{D(D+1)}{2} -1} }  \, .
\eea 
Consider first the case the case $\beta=0$. We have
\bea
 U = S(0,z) = \frac{1-2z-\sqrt{1-4z}}{2z} \Rightarrow 
 g(0,U)=\frac{U}{(1+U)^{D+1}} \, .
\eea 
The function $g(0,U)$ has an unique maximum in $U =\frac{1}{D} $. The curves $g(0,U)$ and $g_b(U)$
intersect for $U=1$, irrespective of $D$. The reader can check that $g_b(U)$ is strictly decreasing for 
$U<1$. The two curves are represented in Fig.~\ref{fig:curves}a.
At the critical point $U=\frac{1}{D}, g_c(0)=g(0,\frac{1}{D}) = \frac{D^D}{(D+1)^{D+1}}$ the function $U$ 
becomes critical, as 
\bea \label{smallbeta}
 ( g -g_c ) \sim ( U -\frac{1}{D} )^2 \Rightarrow U - \frac{1}{D} \sim (g-g_c)^{\frac{1}{2}} 
\Rightarrow E \sim (g-g_c)^{\frac{3}{2}} \; .
\eea 
This is the standard branched polymer phase, obtained in the i.i.d. colored tensor model setting in \cite{Bonzom:2011zz}.

We now slowly turn on $\beta$. The derivative of $S(\b,z)$ w.r.t. $z$ is 
$ \Big{(} \frac{\partial S}{\partial z} \Big{)}_{\beta} = \frac{1}{z}S(\beta-1,z)$. 
The derivatives of the solution $g(\beta,U)$ of the equation \eqref{eq:sys} are 
\bea \label{totdiff}
&& U = S(\beta, z(g,U))\Rightarrow dU = \partial_{\beta} S d\beta + \partial_z S 
\Bigl[ \partial_U z \, dU + \partial_g z\, dg  \Bigr]  \; ,\\
&& \Big{(} \frac{\partial g}{ \partial U}\Big{)}_{\beta}= 
   \frac{ 1 - \partial_z S  \partial_U z  }{\partial_z S  \partial_g z } 
   =  \frac{  1 - S(\beta-1,z) \frac{ D(D+1)-2 - S(\beta,z) D(D-1)   }{D(D+1)  S(\beta,z) \bigl( 1+S(\beta,z) \bigr)  }    }
  {  S(\beta-1,z) \frac{2}{D(D+1)g}  } \; , \crcr
&&  
\Big{(}\frac{\partial g}{\partial \beta} \Big{)}_U = 
-\frac{\p_{\beta}S}{ \p_z S \p_g z } \nonumber \; ,
\eea
where $z= z\bigl( g(\beta, U ),U \bigr)$. The function $g(\beta,U)$ becomes critical for
 $ (\frac{\partial g}{ \partial U})_{\beta}=0$, $i.e.$ at $U=U_c^<(\beta)$ solution of
\bea \label{eq:gcmic1}
1- \partial_z S \Big{|}_{\beta, z=z \bigl( g(\beta, U), U \bigr)} \partial_U z \Big{|}_{ g = g(\beta, U), U } =0 \; ,
\eea 
thus the critical curve $g_c^{<}(\beta)$ is characterized by
\bea \label{eq:gcmic}
 g_c^{<}(\beta) = g\bigl( \beta, U_c^<(\beta) \bigr) \; , \quad z_c^{<}(\beta) =z\bigl(g_c^{<}(\beta) , U_c^<(\beta) \bigr)  \;  ,
\quad U_c^<(\beta) = S \Bigl(\beta, z_c^{<}(\beta)  \Bigr) \; .
\eea 
Equation  \eqref{eq:gcmic1} translates in parametrized form in terms of $z_c^{<}(\beta)$ (dropping the argument $ \beta $ to simplify notations) as  
\bea \label{gc}
 1 - S(\beta-1,z_c^<) \frac{ D(D+1)-2 - S(\beta,z_c^<) D(D-1)   }{D(D+1)  S(\beta,z_c^<) \bigl( 1+S(\beta,z_c^<) \bigr)  }   = 0 \; .
\eea 
Using \eqref{eq:gcmic} , with \eqref{totdiff} and \eqref{eq:gcmic1}, we obtain
\bea\label{eq:dermic}
  \f { d g_c^{<}}{d \beta} =  \Big{(}\f{\p g}{\p \beta} \Big{)}_U\Big{|}_{ \beta, U_c^<(\beta) } = 
-\frac{\partial_{\beta } S }{ \partial_z S  \partial_g z } \Big{|}_{ \beta, U^{<}_c(\beta) } > 0 \; ,
\eea 
thus the critical curve $g_c^{<}(\beta)$ is increasing with $\beta$. As 
$ \big( \frac{\partial^2 g}{ \partial U^2} \big)_{U=U_c} <0$ for all $\b$,  the critical behavior remains that of \eqref{smallbeta}.

However this holds only for small enough $\beta$. Indeed we find that for $\beta>\b_c>1$,
 $g(\beta,U)$ exits the analyticity domain of $S(\beta,z)$ before it can reach its first maximum. 
Some curves  $g(\beta,U)$ for increasing values of $\b$ are represented in Fig.~\ref{fig:curves}b.
\begin{figure}
 \includegraphics[width=5.5cm]{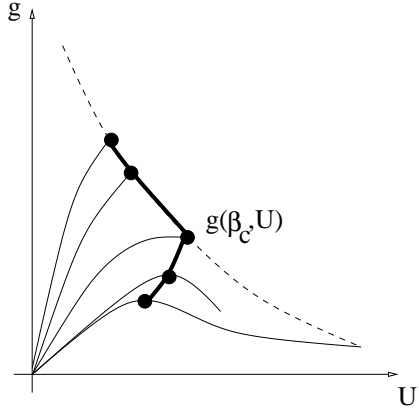}
  \caption{A schematic representation of the phase transition and the critical line $g(\b,U_c(\b))$ (in bold), intersecting the line $g(\b_c,U)$.}
\label{fig:gbg0mare}
\end{figure}

For $\b>\b_c$ both $ \frac{\partial g}{\partial U} $ and $ \frac{\partial U}{\partial g} $ remain finite all the way up to (and including) the boundary $g_b(U)$.
The self consistency equation becomes
\bea
\Big{(} \frac{\partial U}{ \partial g} \Big{)}_{\beta}= 
\frac {  S(\beta-1,z) \frac{2}{D(D+1)g}  } 
{  1 - S(\beta-1,z) \frac{ D(D+1)-2 - S(\beta,z) D(D-1)   }{D(D+1)  S(\beta,z) \bigl( 1+S(\beta,z) \bigr)  }    } \; ,
\eea 
and the denominator stays finite. Close to $z=\frac{1}{4}$, the non-analytic behavior is given by $S(\beta-1,z) \sim (1-4z)^{\beta -1 + \frac{1}{2}}$,
hence
\bea \label{largebeta}
 \Big{(} \frac{\partial U}{ \partial g} \Big{)}_{\text{non-analytic}} \sim (g-g_b)^{\beta -1+\frac{1}{2}} \Rightarrow E_{\text{non-analytic}}
\sim (g-g_b)^{\beta + \frac{3}{2}} \; .
\eea 
Note that, as $\b_c>1$, for $\b>\b_c$ the non-analytic part of the free energy $E$ is always preceded by an analytic part with non-zero
linear and quadratic terms.

It follows that for $\beta>\beta_c$ the critical curve $g^{>}_c(\beta)$ is defined by 
\bea\label{eq:gcmare1}
 g_c^{>}(\beta) \;\text{ solution of }\; z(g, S(\beta,1/4 ) ) = \frac{1}{4} \; , 
\eea
that is $g_c^>(\beta) = g_b \bigl( S(\beta, 1/4) \bigr)$.
Along the critical curve $g^{>}_c(\beta)$  we have
\bea\label{eq:gcmare} 
 U_c^{>}(\beta) = S(\beta, 1/4) \; , \qquad g_c^>(\beta) = g_b \bigl(U_c^{>}(\beta) \bigr) \; , \qquad z_c^{>}(\beta)= \frac{1}{4} \; .
\eea 
The derivative of $g_c^>(\beta)$ computes using the implicit function theorem as
\bea\label{eq:dermare}
 \frac{dg_c^{>}}{d\beta} = - \frac{ \partial_U z \partial_{\beta }S  } { \partial_g z} \; \Big{|}_{ \beta, U_c^{>}(\beta) } > 0 \; ,
\eea 
as $U_c^>(\beta)<1$.

The critical $\beta_c$ corresponds to the point where the two critical curves $g_c^{<}(\beta)$ and $g_c^{>}(\beta)$ meet, that is 
\bea\label{eq:betac}
  g_c^{<}(\beta_c) = g_c^{>}(\beta_c) \;  , \quad U_c^{<}(\beta) = U_c^{>}(\beta) \; , \quad z_c^{<}(\beta) = z_c^{>}(\beta) \; ,
\eea
which translates, using equation \eqref{gc}, \eqref{eq:gcmare} and \eqref{eq:betac}, as the single equation
\bea
 S(\beta_c-1, 1/4 ) = \frac{ D(D+1) S(\beta_c, 1/4) [ 1+ S(\beta_c, 1/4) ]  }{  D(D+1)-2 - S(\beta_c,1/4 ) D(D-1) } \; .
\eea
Numerically we find $\b_c\approx 1.162$ for $D=3$, $\b_c\approx 1.216$ for $D=4$, and  $\b_c\approx 1.134$ for $D=\infty$.
A schematic portrait of the transition is presented in Fig.~\ref{fig:gbg0mare}.

The critical behavior is different at $\b=\b_c$, as at $g=g_b=g_c$ we have at the same time 
$ \Bigl(\frac{\partial g}{\partial U}\Bigr)_{\beta}=0 $, and a non-analytic behavior of $S(\b,z)$ with exponent
 $\b_c+\f12<2$, hence
\bea \label{critbeta}
 ( g -g_c ) \sim ( U - U_c )^{\b_c +\f12} \Rightarrow U - U_c  \sim (g-g_c)^{\frac{1}{\b_c +\f12}} 
\Rightarrow E \sim (g-g_c)^{\frac{\b_c +\f32}{\b_c +\f12}} \; .
\eea 

In conclusion the susceptibility exponent, defined by $E_{\text{non-analytic}} \sim (g-g_c)^{2-\g}$, is 
\be \label{gamma}
\g = \left\{
\begin{array}{ll}
\f12 & \hspace{.5cm} \text{for} \,\,\,\, \b <\b_c \, ,\\
\f{\b_c-\f12}{\b_c+\f12} & \hspace{.5cm} \text{for} \,\,\,\, \b=\b_c \, ,\\
 \f12-\b &\hspace{.5cm}   \text{for} \,\,\,\, \b>\b_c \, .
\end{array} \right.
\ee

We observe that for a large range of values of $\beta$,   $i.e.$ for $\b<\b_c$, universality holds: the critical exponent 
$\g$ is independent of $\b$. 
On the contrary, for $\b\geq \b_c$ we have a one-parameter family of different critical behaviors. A possible interpretation 
of such unusual behavior is that 
for $\b$ sufficiently large the measure term starts behaving as non-local or long-range interaction, for which universality 
is not expected to hold.

\subsection{Order of the phase transition}

Introducing the canonical DT partition function via
\be
Z_{DT} (\k_D,\k_{D-2})= \sum_{N_D} e^{-\k_D N_D} Z_{DT,\text{can}} (N_D,\k_{D-2}) \, ,
\ee
one has that the thermodynamic limit for the DT free energy is given by
\be
F_\infty = \lim_{N_D\to\infty } \f{1}{N_D} \ln Z_{DT,\text{can}} (N_D,\k_{D-2}) \sim -\ln g_c(\b) \, .
\ee
The order of the phase transition around $\b_c$ is then assessed by studying the discontinuity of $g_c(\b)$ or its derivative 
at $\b_c$. Combining eq. \eqref{eq:dermic}, with \eqref{eq:dermare} and using eq. \eqref{eq:gcmic1} we obtain 
\bea
  \frac{dg_c^{>}}{d\beta} (\beta_c)= \f { d g_c^{<}}{d \beta} (\beta_c) \; ,
\eea 
thus the phase transition is higher than first order!

To check the precise order of the phase transition we re-parametrize the self consistency equation by eliminating $U$
\bea
&& H(g,U) =  \frac{ U^{1-\frac{2}{D(D+1) } }   }
{  \bigl( 1+ U \bigr)^{2 \frac{D-1}{D}}  }   g^{\frac{2}{D(D+1)}} \; ,  \qquad 
 S(\beta,z) =  \sum_{q=1}^\infty \frac{1}{q^{\beta} (q+1) } \binom{2q}{q} z^q \; ,\crcr
&&  z = H \bigl(g,S (\beta,z) \bigr) \; , \qquad g( \beta, 0  ) =0 \; ,
\eea 
and denote its solution $g(\beta, z)$. We therefore have
\bea
&&  dz = \p_g H dg + \p_U H \bigl( \p_{\beta} S d\beta + \p_z S dz  \bigr) \; ,\crcr
&& \Bigl( \f{\p g}{ \p z} \Bigr)_{\beta} = \f{ 1 - \p_z S \p_U H }{ \p_g H } \; , \qquad
 \Bigl( \f{\p g }{\p \beta}   \Bigr)_z = - \frac{ \p_U H \p_{\beta} S  }{ \p_g H } \; .
\eea 
The two critical curves are
\bea
 && g^<_c(\beta) = g(\beta, z^<_c(\beta)) \;\text{ with }\; z^<_c(\beta)\; \text{ solution of }\; 1 - \p_z S \p_U H = 0 \; ,\crcr
 && g^>_c(\beta) = g( \beta, \frac{1}{4} )\; \text{ with } \; z^>_c(\beta) = \frac{1}{4}, \;\text{ i.e. }
  g^>_c(\beta) \;\text{ solution of } \;\frac{1}{4} = H(g, S (\beta, 1/4)) \; .
\eea 
and meet at $\beta_c$, when
\bea
z^<_c(\beta_c) = \frac{1}{4} \qquad  g^>_c(\beta_c) = g^<_c(\beta_c) \; .
\eea 
It follows that the derivatives of the critical couplings are 
\bea
 && \frac{dg_c^<}{d\beta} = \Bigl( \f{\p g}{\p \beta}\Bigr)_z = - \frac{ \p_U H \p_{\beta} S  }{ \p_g H } \Big{|}_{ \beta, z_c^<(\beta)  } 
=- \frac{ \p_U H\Bigl( g^<_c(\beta), S \bigl( \beta, z^<_c(\beta) \bigr) \Bigr) \p_{\beta} S \bigl( \beta, z^<_c(\beta)\bigr)  }
{ \p_g H\Bigl( g^<_c(\beta), S \bigl( \beta, z^<_c(\beta) \bigr) \Bigr) } \; ,
\crcr
 && \frac{dg_c^>}{d\beta} = -\frac{ \p_U H \p_{\beta} S  }{ \p_g H} \Big{|}_{ \beta, 1/4  }
=- \frac{ \p_U H\Bigl( g^>_c(\beta), S \bigl( \beta, 1/4 \bigr) \Bigr) \p_{\beta} S \bigl( \beta, 1/4 \bigr)  }
{ \p_g H\Bigl( g^>_c(\beta), S \bigl( \beta, 1/4  \bigr) \Bigr) } \; ,
\eea 
which,  as we already knew, are finite and continuous at the critical point. The new parametrization comes in handy once 
we move to higher order derivatives.

All the derivatives of  $g_c^>$ are finite at $\b=\b_c$ (note that $H(g,U)$ is analytic for $g>0,U>0$, hence its derivatives just
go to constants at $\beta_c, z = \frac{1}{4}$), and the derivatives of $g_c^<$ will differ from them due to terms involving the 
derivatives of  $z_c^<$
\bea
 &&\lim_{\D\b\to 0^+} \Big{(} \frac{d^2g_c^<}{d\beta^2}\Big{|}_{\beta_c -\D\b} 
-   \frac{d^2g_c^>}{d\beta^2}\Big{|}_{\beta_c+\D\b} \Big{)} \sim \frac{dz^<_c}{d\beta} \; .
\eea 

We thus only have to check at which order the derivatives of  $z_c^<$  are non-zero and/or singular.
In order to evaluate $\frac{dz^<_c}{d\beta}  $ we use
\bea
 d\bigl( \p_z S \p_U H \bigr) =0 \Rightarrow \f{dz^<}{d\beta} 
= - \frac{ \p_{\beta z} S \p_U H  -  \p_{gU} H \frac{ \p_{\beta}S }{\p_g H}   + \p_z S  \p_{\beta} S\p_{UU}H    }
{  \p_{zz}S  \p_U H  + (\p_z S)^2 \p_{UU}H  } \, .
\eea 
The singular behavior comes from approaching the convergence radius of $S(\b,z)$, where we have
\bea
 S(\beta, \frac{1}{4} - \Delta z) \approx S(\beta ,\frac{1}{4}) - 4 S(\beta-1, \frac{1}{4}) \Delta z 
+ \frac{\Gamma(-\beta-\frac{1}{2})}{\sqrt{\pi}} 
\Delta z^{\beta+\frac{1}{2}} \, ,
\eea 
from which we deduce that at $\b=\b_c$ the first singularity is in
\bea
 \partial_{zz} S \sim \Delta z^{ \beta_c - \frac{3}{2} } \Rightarrow \f{dz^<}{d\beta} \sim \Delta z^{ \frac{3}{2} -\b_c}   \; .
\eea 

Finally we use the self-consistency equation and the expansion of $S(\b,z)$ to get 
 $\Delta z \sim \Delta \beta^{\frac{1}{\beta_c - \frac{1}{2}}}$,
and
\bea
 \f{dz^<}{d\beta} \sim  \Delta \beta^{ \frac{ \frac{3}{2} - \beta_c}{ \beta_c - \frac{1}{2} } } \xrightarrow{\Delta \beta\to 0} 0
\,\, \Rightarrow  \,\,
  \frac{d^2g_c^<}{d\beta^2}(\b_c) =   \frac{d^2g_c^>}{d\beta^2} (\b_c) \, ,
\eea 
while 
\bea
\lim_{\D\b\to 0^+} \Big{(} \frac{d^3g_c^<}{d\beta^3}\Big{|}_{\beta_c -\D\b} 
-   \frac{d^3g_c^>}{d\beta^3}\Big{|}_{\beta_c+\D\b} \Big{)} 
 \sim \f{d^2 z^<}{d\beta^2} \sim \Delta \beta^{ \frac{ 2-2\beta_c }{\beta_c - \frac{1}{2}}  } \xrightarrow{\Delta \beta\to 0} \infty\, ,
\eea 
thus the transition is third order.

\subsection{Comparison to branched polymers}
\label{Sec:BP}

The $\beta$-dependent behavior we found is reminiscent of that of certain models of branched polymers (BP) \cite{Ambjorn:1985dn,Bialas:1996ya}. 
Such models are defined by the partition function
\be
Z_{BP} (\mu) = \sum_N e^{-\mu N} \bar{Z}_{BP}(N) \, ,
\ee
where
\be
\bar{Z}_{BP}(N) = \sum_{BP_N} \prod_{i=1}^N p(n_i) 
\ee
is the canonical partition function for rooted trees $BP_N$ with $N$ vertices, and $n_i$ is the degree of the vertex $i$.
Making the choice
\be
p(n) = n^{-\alpha} \, ,
\ee
one finds a critical behavior\footnote{Note that as a result of the presence of a root, the partition function for rooted BP is interpreted 
as the derivative of a ``non-rooted'' model.} $Z_{BP} (\mu) \sim (\mu-\mu_c)^{1-\gamma}$, with the susceptibility exponent $\gamma$ given by
\be \label{gammaBP}
\g = \left\{
\begin{array}{ll}
\f12 & \hspace{.5cm} \text{for} \,\,\,\, \a <\a_c \, ,\\
\f{\a_c-2}{\a_c-1}\simeq 0.3237 & \hspace{.5cm} \text{for} \,\,\,\, \a=\a_c=2.4787 \, ,\\
 2-\alpha &\hspace{.5cm}   \text{for} \,\,\,\, \a>\a_c \, .
\end{array} \right.
\ee

Comparing to the DT model we studied here, we see that we get qualitatively the same kind of behavior 
(up to the precise value of the critical coupling) if we identify $\a=\b+3/2$. The extra $3/2$ is coming from 
the asymptotic scaling of the factor $ \frac{1}{q+1} \binom{2q}{q} \sim q^{-3/2}  4^q$, and can be understood 
as the natural entropy factor of the triangulations with trivial measure.
The precise value of the critical coupling differs in the two cases (as for any $D$ we have $\b_c+3/2>\a_c$), 
but this comes as no surprise as critical couplings are generically non-universal quantities.
For example one can adjust its value, without affecting anything else in \eqref{gammaBP}, along the lines of  
\cite{Bialas:1996ya}: introducing an additional weight
 $t$ for the the number of vertices on the last generation of branches of the tree, one finds that the critical point changes 
with $t$, while the
 susceptibility exponent above and below the transition remains unaltered. For that particular modification one finds that
 decreasing $t$ leads to an
 increase in $\a_c$, and that for $\a_c>3$ the value of $\g$ at the critical point remains frozen at $\g=1/2$.

It is very tempting to exploit the apparent connection to BP in order to extract other properties of the model. For example,
 a number of results are 
known about the Hausdorff dimension $d_H$ \cite{Ambjorn:1990wp} and spectral dimension $d_S$ \cite{Jonsson:1997gk,Correia:1997gf} of BP. 
In particular, following  \cite{Correia:1997gf} we could conjecture that, for $\g>0$ (that is for $\b\leq\b_c$ in our model)
\be \label{dBP}
d_H = \f{1}{\g} \, , \,\,\,\, d_S = \f{2}{1+\g} \, ,
\ee
while for $\g<0$ ($i.e.$ $\b>\b_c$)
\be
d_H = \infty \, , \,\,\,\, d_S = 2 \, .
\ee
For $\b<\b_c$ we have the standard BP phase with $d_H=2$, in agreement with the numerical results for DT \cite{Ambjorn:1995dj}. 
For $\b>\b_c$ we enter a new phase, which doesn't seem to have been observed yet in simulations of DT.
From the BP point of view it is understood that for $\b>\b_c$ the dominating trees are actually short bushes, with the
 limit of $\b\to\infty$ eventually 
represented by a bush with all the branches attached to a unique vertex. Given this intuition, and $d_H = \infty$, one could hypothesize 
that the new phase corresponds to the crumpled phase of DT (which at $\b=0$ is only visible at small $\k_{D-2}$). 
However, as the crumpled phase has $\g=-\infty$, such an interpretation can only make sense for $\b\to\infty$.  

Interestingly at $\b=\b_c$ we could get $2<d_H<\infty$, as $\g>0$. One then could possibly tweak the model so that the value of
$\b_c$ at the critical point fixes either $d_H=D$ or $d_S=D$.\footnote{The relations \eqref{dBP} imply that branched polymers can have 
$d_S=d_H$ only for $\g=1$, thus it seems unlikely to be able to fit both dimensions to $D$ simultaneously.} 

However the reader should be warned that it is not at all obvious whether the parallel with BP should be extended to the Hausdorff 
and spectral
dimensions. 
Sharing one critical exponent is of course not a sufficient 
reason to believe that all critical exponents are common. The relation between melon graphs and trees was already made evident in 
\cite{Bonzom:2011zz}, where 
it was used as an exact bijection for the combinatorial counting. On the other hand, and most importantly, the notion of neighborhood 
in a triangulation 
seems to be very poorly represented by the 
abstract trees associated to melons. Nevertheless it is intriguing that a strong relation between DT (for $\b=0$) in the weak coupling 
phase and BP is 
supported both by numerical 
simulations \cite{Ambjorn:1995dj} (supporting for example  $d_H=1/\g=2$) and by theoretical arguments \cite{Ambjorn:1996ny}. 
Whether or not the link can be clarified further in the context of the colored tensor models is an open issue.

\section{Final discussion and outlook}
\label{Sec:concl}

In this paper we have introduced and solved a new class of colored tensor models. The new models, dubbed dually weighted colored 
tensor models, 
allow to associate arbitrary weights to the faces of the graphs, $i.e.$ to the bones of the dual triangulation.
Choosing the weights as in \eqref{betaC}, we have explicitly computed the susceptibility exponent of the model, and showed that in 
the continuum limit 
the model admits two phases separated by a third order phase transition.
The results are reminiscent of certain models of branched polymers, with which the susceptibility exponent shares the same
qualitative behavior. However, we have not computed the 
Hausdorff dimension for our model, and we have at the moment no reason to believe that it agrees with that of the BP.

We believe that our results are very important for the DT approach to quantum gravity. First of all, this is to our knowledge the 
first time that a phase 
transition in the continuum limit is accessed by analytical means in dimensions higher than two. Furthermore, it is the first time that a 
\emph{third-order} phase transition is
 observed in DT, which opens up the possibility of obtaining a continuum limit with $2<d_H<\infty$, without the causality condition
 employed in CDT
 \cite{Ambjorn:2010ce,Ambjorn:2005qt,Ambjorn:2011ph,Ambjorn:2011cg}. 
The link to the numerical results of \cite{Laiho:2011ya} is unclear, and deserves further exploration, either by pushing the 
simulations to 
larger $\k_{D-2}$ and larger volumes, or by trying to extend the analytical tools to finite $\k_{D-2}$.

It should also be mentioned that the BP models we have discussed  in Sec.~\ref{Sec:BP} have been mapped and generalized to a
 balls-in-boxes models
 \cite{Bialas:1996eh,Bialas:1997qs,Bialas:1998ci},
which has been interpreted as a mean field model of DT. Indeed the model successfully reproduces the first-order transition
 (at $\b=0$ and $\k_{D-2}=\k_{D-2}^c$) between a crumpled and a BP phase, as in the DT simulations \cite{Bialas:1996wu,deBakker:1996zx}, 
and in the $\{\b,\k_{D-2}\}$ plane it shows a phase diagram which is reminiscent of the one found in \cite{Bilke:1998vj} (and more recently 
revisited in \cite{Laiho:2011ya}). In the light of that, the new phase we observed might extend at finite $\k_{D-2}$ into a phase similar 
to the ``condensed phase'' of \cite{Bialas:1998ci} or the ``crinkled phase'' of \cite{Bilke:1998vj}.
Indeed our result of a negative and $\b$-dependent exponent $\g$ seems compatible with the findings of  \cite{Bilke:1998vj}.

In this work we have concentrated on a specific choice of weights, but the solution provided in Sec.~\ref{Sec:largeN} can be used 
for any other choice  of matrix $C$. For example, we could consider
\bea \label{expC}
 \text{Tr}[ (C^TC)^q] = N q^{-\b} e^{\m q^n} \; .
\eea
The case $n\ge 1$ is not very interesting. The modification either shifts the critical point ($n=1$) 
or makes the series in \eqref{selfceq} always convergent (resp. divergent)  for $n>1$, $\m<0$ (resp. $n>1$, $\m >0$).
An interesting choice from the DT point of view is $n=-1$, corresponding to the DT amplitude
\be\label{hdC}
A(\cG) = e^{\kappa_{D-2}N_{D-2}\, -\, \kappa_D N_D+\sum_i \f{\m}{q_i}}   \prod_{i} q_i^{-\b} \, .
\ee
Such amplitude corresponds to the addition of an $R^2$ term to the Regge action \cite{Ambjorn:1992aw}.
For $n<0$ \eqref{expC} the asymptotic of the series in \eqref{selfceq} are unaffected, and the large-$N$ limit of the model would be 
the same as the one we have found for $\m=0$. This is in agreement with the findings of \cite{Ambjorn:1992aw}, according to which 
the inclusion of 
higher-derivative terms does not significantly affect the phase diagram of DT.
Finally, for  $0<n<1$ we have $\g=-\infty$ for $\b>\b_c$. In such case the similarity between the $\b>\b_c$ phase and the 
crumpled phase of DT 
seems stronger than what we have outlined in Sec.~\ref{Sec:BP}. It would be interesting to study this model further and check 
whether the phase 
transition remains third-order or it becomes second or first order .

Another option would be to modify the model by choosing $C$ such that at the critical point $g_c$ the first $m>1$ derivatives 
of the coupling 
become zero, rather than just the first one. In all likelihood this would still relate to some type of branched polymers, like 
the multicritical 
BP studied in \cite{Ambjorn:1990wp}.

The most important open problems in our view are to study whether the connection to branched polymers extends to other critical
 exponents, and in 
particular to the effective dimension of the triangulations. Even more important would be to access finite $\k_{D-2}$, maybe 
via some sort of double scaling limit.

\section*{Acknowledgements}

Research at Perimeter Institute is supported by the Government of Canada through Industry Canada and by the Province of Ontario 
through the Ministry of Research and Innovation.

\end{document}